\documentclass[11pt,letterpaper]{article}
\usepackage[margin=1in]{geometry}
\usepackage[utf8]{inputenc}
\usepackage{subcaption}

\title{Can laypeople predict the replicability of social science studies without expert intervention: an exploratory study}
\usepackage[numbers]{natbib}
\usepackage{graphicx}
\usepackage{multirow}
\usepackage{authblk}

\author[1]{Juntao Wang%\thanks{juntaowang@g.harvard.edu}
}
\author[2]{Jonathan Lei%\thanks{jonathanlei2005@gmail.com}
}
\author[3,4]{Anna Dreber%\thanks{C.C@university.edu}
}
\author[5]{Michael Gordon%\thanks{D.D@university.edu}
}
\author[3]{Magnus Johannesson%\thanks{E.E@university.edu}
}
\author[5]{Thomas Pfeiffer%\thanks{E.E@university.edu}
}
\author[1]{Yiling Chen%\thanks{E.E@university.edu}
}
\affil[1]{John A. Paulson School of Engineering and Applied Sciences, Harvard University, Cambridge, MA, USA}
\affil[2]{Acton-Boxborough Regional High School}
\affil[3]{Department of Economics, Stockholm School of Economics, Stockholm, Sweden}
\affil[4]{Department of Economics, University of Innsbruck, Innsbruck, Austria}
\affil[5]{New Zealand Institute for Advanced Study, Massey University, Auckland, New Zealand}

\usepackage{setspace}
\usepackage{color}
\newcommand{\blue}[1]{\begin{color}{black}#1\end{color}}

\begin{document}

\maketitle

\begin{abstract}
The low replication rate of published studies has long concerned the social science community, making understanding the replicability a critical problem. Several studies have shown that relevant research communities can make predictions about the replicability of individual studies with above-chance accuracy. Follow-up work further indicates that laypeople can also achieve above-chance accuracy in predicting replicability when experts interpret the studies into short descriptions that are more accessible for laypeople. The involvement of scarce expert resources may make these methods expensive from financial and time perspectives. 
In this work, we explored whether laypeople can predict the replicability of social science studies without expert intervention. We presented laypeople with raw materials truncated from published social science papers and elicited their answers to questions related to the paper. Our results suggested that laypeople were engaged in this technical task, providing reasonable and self-contained answers. The majority of them also demonstrated a good understanding of the material. However, the solicited information had limited predictive power on the actual replication outcomes. We further discuss several lessons we learned compared to the approach with expert intervention to inspire future works. 
   
\end{abstract}

\begin{spacing}{1}

\section{Introduction}

The replicability of scientific studies has played a critical role in the development of science~\cite{sep-scientific-reproducibility,schmidt2016shall}. Scientific results that fail to replicate will misguide the research advancement, and impair the credibility of the research community. Concerns of the replicability of social science studies have been raised long ago~\cite{ioannidis2005most,ioannidis2013s,maniadis2014one}, followed by several large-scale replication projects conducted to systematically examine the replicability of published studies across various fields of social science. Four notable such projects conducted in the recent decade are the Reproducibility Project: Psychology (RPP)~\cite{open2015estimating}; the Social Science Replication Project (SSRP)~\cite{camerer2018evaluating}; the Many Labs 2 Project (ML2)~\cite{klein2018many}; and the Experimental Economics Replication Project (EERP)~\cite{camerer2016evaluating}. 
They found a low replication rate ranging from 36\% (RPP) to 62\% (SSRP), which further heated the debate of the replication crisis in social science~\cite{baker20161,christensen2016transparency,fanelli2018opinion}. 

Besides these replication projects, efforts have also been made to develop scientific methods to forecast the replication probability of social science studies, aiming to provide a fast and more economical alternative to indicate the reliability of studies. Accompanying with the four replication projects, four forecasting projects were conducted to explore the potential of using the collective intelligence of the research community to predict the replication probabilities of social science studies~\cite{dreber2015using,camerer2018evaluating,forsell2019predicting,camerer2016evaluating}. In these forecasting projects, hundreds of experts were recruited from relevant research communities to predict the success probabilities of these replication experiments via either surveys or prediction markets. All these projects achieved above-chance prediction accuracy, ranging from 58\% to 86\%, with an average of 66\% for surveys and 73\% for prediction markets~\cite{gordon2021predicting}, showing the effectiveness of this approach. Inspired by the success of this crowd forecasting approach, the Defense Advanced Research Projects Agency (DARPA)'s program ``Systematizing Confidence in Open Research and Evidence (SCORE)'' further used this approach to generate confidence scores for thousands of social science studies and investigate the replicability differences across fields~\cite{gordon2020replication}.

However, expert resources are usually scarce and expensive. The development of online crowdsourcing markets like the Amazon Mechanical Turk platform has enabled us to access laypeople’s intelligence much more flexibly, inexpressively, and scalable than accessing expert resources. Using laypeople to make collective forecasts has been proven to be surprisingly accurate in various applications, such as predicting the outcomes of geopolitics, economics,
or sports events~\cite{mellers2014psychological,goldstein2014wisdom}. This phenomenon is referred to as the \emph{wisdom of crowds}, which has been extensively researched~\cite{prelec2017solution}, with the earliest research dating back to a hundred years ago~\cite{galton1907vox}. Recently, Hoogeveen et al.~\cite{hoogeveen2020laypeople} have conducted the pioneering work of using laypeople's wisdom to predict the replicability of social science studies. 
They presented the participants with the materials of 27 selected studies from SSRP and ML2 and asked them to provide a replication prediction for each study. They achieved an accuracy of 59\% when presenting the participants with a short description of the study and 67\% when additionally presenting with the Bayes factor and its verbal interpretation of the study. However, they still required experts to compose these short descriptions that are comprehensible to laypeople, which may be a potential bottleneck to the scalability of their
approach.

In this work, we explored the potential of using laypeople to make predictions about the replicability of social science studies without expert intervention. We investigated to what extent we could elicit useful information when we presented laypeople with raw materials truncated from the published papers of the studies. In particular, we had three objectives: i) evaluating laypeople's engagement in such technical tasks, ii) knowing their perceptions about social science studies from the perspectives of the surprisingness of the findings and the accessibility of the raw materials, and iii) predicting the replicability of social science studies using the solicited information. In the following, we provide a methods section introducing details about our experiment design and data collection process. This is followed by a results section describing our findings and related statistical analysis and a discussion section discussing our results in light of Hoogeveen et al.'s results.

\section{Methods}

\subsection{Materials}
We selected 89 studies out of the 97 published studies investigated in RPP. These 89 studies have both a known replication outcome and an abstract section. We released the surveys about these studies using Human Intelligence Task (HIT) on the Amazon Mechanical Turk (Mturk) platform. Each HIT contained two surveys about two studies uniformly randomly selected from the 89 studies and an additional exit survey about participant's demographic information and user experience. 

\subsubsection{Presentation of studies and survey questions}
In each survey, the reading material and the survey questions were presented using two web pages to reduce participants' cognitive burden. The participants could proceed to the second page only after completing the questions on the first page. The first page presented the title and the abstract of the study and contained four selection and rating questions. Each numerical rating was associated with a brief description of this rating. The four questions are listed below. 
\begin{itemize}
    \item (Q1) Please select a single sentence in the abstract that best describes the main findings/claims of the paper.
    \item (Q2) In the abstract, how many phrases or terms are NOT familiar to you?  Rate between 1 to 4.\footnote{1=``0''; 2=``1 or 2''; 3=``3 or 4''; 4=``5 or more''.}
    \item (Q3) How much do you understand the main findings/claims of the study after reading the abstract? Rate between 1 to 4.\footnote{1=``It is very clear to me what the study has found'';
    	2=``I have some general ideas about what the study has found but still find some places unclear'';
    	3=``I am not so sure about what the study has found, but I can make a rough guess'';
    	4=``I have no idea what the study attempts to achieve''.}
    \item (Q4) Do you find the main findings/claims surprising? Rate between 1 to 4.\footnote{1=``Completely unsurprising''; 2=``Somewhat unsurprising'';3="Somewhat surprising"; 4=``Completely surprising''}
\end{itemize}
Participants were also asked to enumerate some unfamiliar terms and phrases and to describe the main findings in their own language in text boxes. 

The second page presented the section of one of the experiments in the study.\footnote{RPP ran replication experiments on the last experiment of each study. Therefore, we aimed to present the section of the last experiment of each study. However, in some studies, the methods of the last experiment referred to those of previous experiments. In these cases, we presented the participants with the section of the experiment referred to.} Our goal was to provide the participants with a taste of how the experiment was designed and conducted to support the study's findings. We excluded the results subsection from the presented material for two reasons. First, we wanted to keep a reasonable length of the reading material for laypeople. Second, the result subsection usually contains many statistical terminologies, which are not accessible to laypeople. The reading material was followed by five selection and rating questions listed below. 
\begin{itemize}
    \item (Q5) Please select a single sentence that best describes what the experiment/study does. Such a sentence usually appears in a paragraph (if exists) before the "Method" section. If you think that there is no such a sentence, please check the box below.
    \item (Q6) Please select three sentences that best describe the most important steps of the experiment/study. %More info: A "step" decribes what the authors do or what the participants are asked to do. The most important steps may appear in different sections of the material. (If there are multiple experiments/ studies/ trials/ treatments, please select sentences only for the first experiments/ studies/ trials/ treatments mentioned.)
    \item (Q7) How many participants were recruited in the experimental study?
    \item (Q8) If the same type of experiments are re-performed, what probability do you assign that the findings presented in the abstract will be observed? Make your best prediction from 0\% (not likely at all) to 100\% (with certainty the same finding).
    \item (Q9)  How do you feel that the findings/claims of the paper may hold in other scenarios besides the scenario tested in the above experiment? Rate between 1 to 3.\footnote{1=``The claims/findings will likely NOT hold in other scenarios''; 2=``The claims/findings may partially hold in some similar scenarios''; 3=``The claims/findings will hold to be true in many other similar scenarios''.}
\end{itemize}

The materials of the studies were presented in the same format as they were presented in the web version (Html full-text version) on sagepub.com and ebscohost.com. 
%In rating questions, each rate was associated with a brief description of this rate. 
In sentence selection questions, the participant could directly click to select the sentence in the given reading material. Brief instructions and examples were given to help participants understand what type of sentences were good selections for each question.  

\subsubsection{Design of survey questions}
The survey questions Q1 to Q9 were designed to explore three dimensions of information we wanted to collect from participants: participants' engagement in our HITs, participants' perception about the studies, and participants' predictions about the studies' replicability.

\paragraph{Participants' engagement:} To laypeople, reading social science studies and answering related questions might be a tedious and non-trivial task. Therefore, we designed the sentence selection questions Q1, Q5, Q6, and the factual question Q7 to help us evaluate participants' engagement in our HITs. We compared the responses from different participants in the sentence selections, Q1, Q5, and Q6, to evaluate whether participants provided random answers. Q7 asked about the number of participants in the given study. We compared the responses to the correct answer to evaluate whether participants answered this question authentically.
Moreover, Q2 and Q3 are both related to the accessibility of the study, and Q8 and Q9 are related to the replicability of the study. Intuitively, if participants completed the surveys in good faith, the answers to each pair of questions should demonstrate positive correlations to some extent. 

\paragraph{Participants' perceptions about the studies:} We designed Q2, Q3, and Q4 to evaluate participants' perceptions of the studies from the perspectives of accessibility and surprisingness. We wanted to investigate how these perceptions correlate with participants' replication predictions. Q2 and Q3 both aimed to evaluate the accessibility of the studies to laypeople. While Q2 asked participants to report the number of unfamiliar terms and phrases in the abstract, Q3  asked participants to rate their understanding of the abstract directly. Q4 asked participants to rate the surprisingness of the main claims/findings presented in the abstracts of the studies.

\paragraph{Participants' replication predictions about the studies:} Q8 and Q9 were designed to elicit participants' predictions about the replicability of the studies. While Q8 asked concretely about a replication probability, Q9 asked about the generalizability of the main claims/findings.

\subsection{Procedure and incentive} 
We conducted the surveys using HITs on the Amazon Mturk platform. Mturk workers could see a preview page of our experiments and then determine whether to take the HITs. The preview page presented the motivations and the purpose of our experiments. It stated that the HIT was about reading social science papers and answering related questions about the accessibility, the plausibility, and the replicability about the studies. The preview page also stated that the estimated completion time of the HIT was 30 minutes, with a fixed \$6 compensation upon completion. 
After the workers accepted the HITs, we showed an instruction page describing that the HIT consisted of reading materials about two published social science studies and answering about ten questions. Meanwhile, in this page, we also stated that ``Your good-faith effort to understand the paper materials and answer the questions is crucial to our experiment and to the development of modern social sciences. We appreciate your contributions!'' 
Participants could move forward only if they checked the box ``I will put forward my good-faith effort in completing the task.'' The next page showed the consent of participation. After that, participants were presented with two surveys about two different studies uniformly randomly selected from the 89 RPP studies, followed by an exit survey that ended the whole HIT. 

Participants could quit the HITs at any time during their participation, but if they did not complete the entire HITs, they would not receive the fixed \$6 payment. If a participant completed a HIT within 10 minutes, they would be blocked from taking more HITs during the next 12 hours. They would be shown a clear message that they were blocked for 12 hours, but the reason was not given to them.

\subsection{Participants}
Participants were recruited from the Amazon Mturk platform. Each participant could take at most 3 HITs per day and 20 HITs in total. Each completed HIT was paid with a fixed \$6 upon completion. 405 Mturk workers completed at least one entire HIT. The median HITs completed by participants is 2 (M=5.50, SD=6.56). According to the exit survey, among the 405 participants,  1.48\% had a Ph.D. or equivalent degree, 10.62\% had a master's degree or were pursuing a Ph.D. degree, 45.19\% had a Bachelor's degree, and 42.72\% were undergraduate students or below. Only 3.21\% indicated that they had previously heard about RPP or similar replication projects.

\section{Results}
We received 2229 complete HITs, corresponding to 4458 complete survey responses for individual papers. Each of the 89 papers received either 50 or 51 responses with a mean of 50.09. The median time spent on a single HIT was 29 minutes (M=35, SD=19). The median of the HIT experience ratings, which ranges from 1 (poor) to 5 (excellent), is 4 (M=3.84, SD=0.92).

\subsection{Participants' Engagement}
We analyzed the responses from the questions related to participants' engagement and observed that the the participants i) correctly answered the factual question Q7, ii) formed consensuses on the sentence selections, and iii) demonstrated anticipated correlation in correlated questions. 

In the factual question Q7,  4089 (91.72\%) out of 4458 responses correctly answered the number of participants in the given study. The majority was correct on 87 out of 89 studies. These results suggested that the participants answered this factual question authentically, and laypeople can identify the number of participants given the corresponding experiment materials truncated from the social science papers. 

In the sentence selection questions Q1 (main finding in the abstract), Q5 (what the study does), and Q6 (most important steps), we observed a salient concentration in the selection, differing from the pattern generated by random selection. Here we presented our detailed analysis of Q1 and Q5 and we observed similar results on Q6. Q1 asked participants to select the main claim sentence in the abstract section, while Q5 asked participants to select a single sentence in the method section that best summarizes what the experiment does. We found that in Q1, out of the 50 responses received on each of the 89 studies, the top 5 most frequently selected sentences on each paper were selected by participants 26.2 (SD=6.4), 14.6 (SD=5.46), 5.2 (SD=2.48),  2.6 (SD=1.62), and 1.2 (SD=0.98) times on average, respectively. In comparison, we simulated the scenario that the participants uniformly randomly selected a sentence in the abstract. Consequently, the top 5 most frequently selected sentences of each paper would be selected 7.26 (SD=1.29), 7.23 (SD=1.27), 7.10 (SD=1.24), 7.04 (SD=1.23), and 6.96 (SD=1.2) times on average, significantly differing from the pattern we observed. \blue{A $\chi^2$-test rejected the null hypothesis that the two distributions were equal with $p<0.0001$.}
In Q5, 28.24\% of the 50 responses per study indicated that there was no single sentence summarizing the experiment method. These responses were associated with specific studies. On average, each study received 14.15 (SD=14.42) such responses, while the median was only 5. The top 25\% (22) studies received 62.75\% of these responses. Meanwhile, among the responses that selected a single sentence, the top 5 most frequently selected sentences in each study were selected 23 (SD=10.14), 7.6 (SD=4.22),  2.8 (SD=0.98), 2.2 (SD=0.98), and 1.4 (SD=0.6) times on average, respectively. In comparison, our simulation showed that the most frequently selected sentence would be selected only 1.43 (SD=0.51) times if participants uniformly randomly selected a sentence in Q5. \blue{A $\chi^2$-test rejected the null hypothesis that the two distributions were equal with $p<0.0001$.}

We further investigated the locations of these selected sentences to examine whether participants tended to select the first or the last sentence regardless of the context. We found that both questions had a considerable number of responses selecting a sentence in the middle, and the distributions of the locations of the selected sentence in the two questions differ significantly. 
In particular, in Q1, 12.1\% of the responses selected the first sentence, 50.98\% selected a sentence in the middle, and 36.95\% selected the last sentence. In Q5, 45.14\% of the responses selected the first sentence, 49.98\% selected a sentence in the middle, and only 4.88\% selected the last sentence. These results suggested that participants did not select the sentence purely based on the sentence location regardless of the context.

We also observed a moderate negative correlation  
%using Kendall's Tau test ($\tau=-0.45$, $p<0.0001$) 
in Spearman's correlation test ($\rho = -0.50$, $p<0.0001$)
between participants' ratings of the number of unfamiliar terms and phrases in the abstract (Q2) and their ratings of the accessibility of the abstract (Q3). 
This result followed the intuition that the more unfamiliar terms and phrases in the material, the harder it was to understand the material. 
We also observed a strong positive correlation 
%using Kendall's Tau test ($\tau=0.57$, $p<0.0001$) 
in Spearman's correlation test ($\rho = 0.68$, $p<0.0001$) between participants' ratings of the generalizability of the results of the given study (Q9) and their replication predictions (Q8). These results suggested that the participants' responses were self-contained.

Overall, these results suggested that the participants in our experiments completed the surveys with effort and good faith. They also suggested that laypeople can read published social science papers, understand the materials to some extent, and answer related questions authentically.

\subsection{Participants' perceptions about studies}

We focused on the perceptions of the accessibility (Q3) and the surprisingness (Q4) of the studies. For the accessibility, most responses indicated that they either clearly understood or had a general idea about the main claims/findings in the abstract. In particular, 26.02\% of the responses had a rating of 4 in Q3, referring to that the participant thought that they clearly understood the main claims. 40.51\% had a rating of 3, indicating that the participant had a general idea about the main claims. 23.62\% gave a rating of 2, meaning that the participant could make a rough guess, and the rest, 9.85\%, gave a rating of 1, indicating that the participant had no idea about the main claims/findings. Each study's mean accessibility rating was concentrated around 3 (M=2.83, SD=0.54, Median=2.72).  

For the surprisingness, most responses found the main claims/findings unsurprising. In particular, 23.53\% gave a rating 1 of completely unsurprising, 43.16\% gave a rating 2 of somewhat unsurprising, 27.16\% gave a rating 3 of somewhat surprising, and the rest, 6.15\%, gave a rating 4 of completely surprising. Each study's mean rating of surprisingness was concentrated around rating 2 (M=2.16, SD=0.30, Median=2.18). There existed a weak negative correlation between participants' surprisingness and accessibility ratings 
in Spearman's correlation test ($\rho = -0.22$, $p<0.0001$).
%in Kendall's tau test $(\tau = -0.19, p<0.0001)$.

\subsection{Forecasting replicability}

\subsubsection{Participants' prediction accuracy}

Among the 89 RPP studies, 36 studies replicated successfully, resulting in a replication rate of 40.45\%.
We used the participants' mean replication prediction from Q8 as the final prediction of the replicability of each study. We observed a salient overestimation of the replication probability in laypeople's predictions. The median of these 89 final predictions was 0.69 with a minimum of 0.58, greater than 0.5 (M=0.69, SD=0.05, Max=0.81). If we threshold final predictions at 0.5, all these final predictions forecasted that the corresponding study could replicate successfully,  resulting in an accuracy score of 40.45\%, lower than random guesses. At the response level, only 595 out of 4458 (13\%) responses gave a replication probability lower than 0.5 with a median of 0.71 (M=0.69, SD=0.19, Min=0.00, Max=1.00).\footnote{We provided participants a sliding bar ranging from 0 to 100 (\%) with step size 1 (\%) to indicate their prediction. The sliding button was initialized at the center position 50. Participants were allowed to submit their answers only if a movement of the sliding button was detected. 2.4\% responses predicted exactly 50 (\%).} 
These results suggested that laypeople tend to believe that a published social science study can replicate successfully. 
The overestimation of social science studies' replicability has also been found in other studies with participants recruited from either the research community~\cite{gordon2020replication,gordon2021predicting} or the laypeople~\cite{hoogeveen2020laypeople}.
To remove the effect of overestimation on the accuracy, we investigated the discriminatory power of the participants' mean predictions and conducted a rank correlation test. 
However, Spearman's correlation test showed no significant rank correlation ($\rho=-0.18$, $p=0.0764$) 
%in Kendall's tau test ($\tau=-0.1552, p=0.0764$).
between the mean predictions and the replication outcomes.
This result suggested that laypeople's replication predictions collected in our experiments contained very limited signals about whether a social study can replicate or not.

We further investigated whether the participants' predictions and their prediction accuracy were influenced by the  accessibility of the studies.
Table~\ref{table_understand} shows the statistics of the responses with different accessibility ratings.
We observed that the mean replication prediction increased with the accessibility rating. In fact, there existed a weak positive Spearman's rank correlation between the reported accessibility and the replication prediction at the response level 
%($\tau=0.17, p<0.0001$) 
($\rho=0.22$, $p<0.0001$)
and a moderate positive correlation between them at the study level
($\rho=0.51$, $p<0.0001$).
%($\tau=0.36, p<0.0001$). 
In contrast, the actual replication rate decreased with the reported accessibility. There existed a very weak negative rank correlation between the reported accessibility rating and the actual replication outcome at the response level 
($\rho=-0.11$, $p<0.0001$),
%($\tau=-0.10, p<0.0001$),
and no significant correlation at the study level
($\rho=-0.18$, $p=0.08$).
%($\tau=-0.15,p=0.08$). 
Meanwhile, at each of the four accessibility levels, there was no discriminatory power found between the participants' replication predictions and the actual replication outcomes, as the 
%Kendall's tau coefficients 
Spearman's rank correlation coefficients
were all close to zero with a p-value greater than 0.10 (last row, Table~\ref{table_understand}). 
These results suggested that the participants tended to give a higher replication prediction to the more accessible studies. However, the reported accessibility of the studies had poor discriminatory power (a weak negative correlation) in predicting the replication outcome. This might explain why the participants' replication predictions also had poor predictive power. 

\begin{table}[t]
\begin{center}
\begin{tabular}{ |c|c|c|c|c| } 
 \hline
 Accessibility & 1 & 2 & 3 & 4 \\ 
 \hline
 \# responses & 439 & 1053 & 1806 & 1160 \\
 Mean replication prediction & 0.61 (0.21) & 0.65 (0.19) & 0.70 (0.17) & 0.74 (0.17) \\
 Actual replication rate & 0.52 (0.50) & 0.44 (0.50) & 0.41 (0.49) & 0.32 (0.47) \\
 %Kendall's tau & 0.02 (p=0.61) & -0.03 (p=0.24) & -0.02 (p=0.25) & -0.02 (p=0.43)\\
 Spearman's $\rho$ & 0.02 (p=0.61) & -0.04 (p=0.24) & -0.03 (p=0.25) & -0.02 (p=0.43)\\
 
 \hline
\end{tabular}
\caption{Statistics of the responses with different accessibility ratings. Spearman's rank correlation coefficient $\rho$ is calculated between participants' replication predictions and actual replication outcomes.%Kendall's tau is calculated between participants' replication predictions and actual replication outcomes.
}
\label{table_understand}
\end{center}
\end{table}

We also observed a similar but smaller correlation between the participants' self-rated surprisingness and the replication prediction performance. The participants' self-rated surprisingness had a very weak negative Spearman's rank correlation to their replication predictions at the response level 
($\rho=-0.15$, $p<0.0001$),
%($\tau=-0.12, p<0.0001$) 
and a weak negative rank correlation at the study level 
($\rho=-0.32$, $p=0.002$). 
%($\tau=-0.21, p<0.0001$). 
This suggested that the participants tended to give a lower replication prediction when they found the main claims/findings in the abstract surprising to them. 
In contrast, the self-rated surprisingness had a very weak positive rank correlation to the actual replication outcome at the response level
($\rho=0.07$, $p<0.0001$)
%($\tau=0.07, p<0.0001$)
and no significant correlation at the study level 
($\rho=-0.19$, $p=0.08$).
%($\tau=0.16, p=0.07$).

\subsection{Forecasting using machine learning}
Machine learning is a technique to learn a pattern from historical data to make predictions on new coming data. Machine learning has been applied to predict the replicability of social science studies in various settings~\cite{altmejd2019predicting,yang2020estimating,salsabil2018study}. 
We investigated whether we can use machine learning to improve the prediction accuracy of people's predictions. We used the responses collected on the 89 RPP studies as our dataset to evaluate the machine learning approach. We divided these 89 studies into a training set and a validation set using the cross-validation method. This setup simulated the situation where we have access to participants' historical prediction data to help us make final predictions.

As we only have a limited 89 samples, we use only three features to predict the replicability of each study: the mean replication prediction, the mean accessibility, and the mean surprisingness received by each study. We use the classic logistic regression as the classifier to avoid over-fitting the data. To investigate the predictive power of each feature, we also evaluated the performance of using every single feature to make forecasts. We focused on two accuracy metrics, the accuracy score and the AUC-ROC~\cite{davis2006relationship}. The latter is a common accuracy metric used in the machine learning community to evaluate the discriminatory power of predictions. We ran 2000 times 5-fold cross-validation on the 89 studies and collected 10000 accuracy scores and AUC-ROC on both the training and validation sets.

\begin{table}[t]
\begin{center}
\begin{tabular}{ |c|c|c|c|c| } 
 \hline
 \multirow{2}{*}{Features} & \multicolumn{2}{c|}{Training} & \multicolumn{2}{c|}{Validation}\\ 
 \cline{2-5}
  & Accuracy & AUC-ROC & Accuracy & AUC-ROC \\
 \hline
 Replication prediction (Q8) & 
 0.62 [0.59, 0.65] & 0.61 [0.59, 0.66] & 0.61 [0.5, 0.72] & 0.60 [0.43, 0.69]
 \\ %\hline
 
 Accessibility (Q3) & 
 0.60 [0.58, 0.62] & 0.61 [0.59, 0.63] & 0.60 [0.53, 0.72] & 0.61 [0.53, 0.71]
 \\%\hline
 
 Surprisingness (Q4) & 
 0.62 [0.59, 0.65] & 0.61 [0.58, 0.66] & 0.62 [0.44, 0.72] & 0.60 [0.40, 0.75]
 \\%\hline

 All (Q3, Q4, Q8) & 
 0.64 [0.61, 0.66] & 0.65 [0.64, 0.68] & 0.63 [0.50, 0.72] & 0.60 [0.44, 0.73]
 \\
 \hline
 
\end{tabular}
\caption{Mean accuracy scores and AUC-ROCs on the training set and validation set when the mean replication prediction (Q8), mean accessibility (Q3) and mean surprisingness (Q4) and all of them are used as features respectively. Brackets show the 95\% confidence intervals of the corresponding values.}
\label{table_ml}
\end{center}
\end{table}

Table~\ref{table_ml} shows the mean accuracy and the mean AUC-ROC and their confidence intervals on both the training set and the validation set, when the mean replication prediction (Q8), the mean accessibility (Q3), and the mean surprisingness (Q4) and all of them were used as learning features, respectively. 
All four sets of features showed similar prediction performance in the accuracy score and AUC-ROC. We observed an improvement in the accuracy score (around 0.60) compared to the participants' raw predictions (0.41). This result demonstrated the potential of using machine learning to correct the bias in laypeople's replication prediction data via learning from historical data.  
However, this improvement was limited, as there was no significant difference in the accuracy score from always predicting that the study could not replicate (which obtained an accuracy score of 0.59 on the 89 RPP studies). Meanwhile, these machine learning predictions achieved an AUC-ROC around 0.6., better than random guesses (AUC-ROC=0.5). This improvement was significant ($p<0.0001$) when the classifier used the mean accessibility rating as the only feature to predict the replicability.
We also observed no improvement in using all features together to predict the replicability compared to using a single feature. This might be because these three features turned out to be correlated with each other, and each feature had limited discriminatory power. %Overall, using machine learning had the potential in improving the prediction accuracy, but the improvement 

\section{Discussion}
In this work, we explored whether laypeople can predict the replicability of social science studies without expert intervention. We carefully designed surveys and collected responses via releasing HITs on the Amazon Mturk platform. Our experiments revealed several interesting findings. 

First, Amazon Mturk workers engaged in our very technical HITs, which involved reading raw material truncated from published social science papers and answering related questions. They devoted considerable time and effort to the HITs and provided reasonable and self-contained answers. This showed the potential of using Amazon Mturk workers to extract information from social science papers that might be difficult to extract via a pure machine approach. 

Second, we found that these social science studies in the RPP projects were accessible to laypeople to some extent, as most responses indicated that they either had a general idea about or clearly understood the main findings of the studies. Participants also formed consensuses about the main sentences that summarized the abstract and the experimental method and that described the main experimental steps.

Third, we found that laypeople's replication predictions or perceptions about the studies in our experiments had limited predictive power in predicting actual replication outcomes. Without expert intervention, laypeople demonstrated a prediction accuracy (40\%) lower than chance in our experiments. In contrast, both the researcher forecasters and laypeople with expert intervention achieved above-chance prediction accuracy in similar survey-based experiments.
Dreber et al.~\cite{dreber2015using} reported a prediction accuracy of 58\% achieved by researcher forecasters on the same RPP paper set. Hoogeveen et al.~\cite{hoogeveen2020laypeople} reported prediction accuracy of 59\% and 67\% achieved by laypeople with varied conditions on 27 selected social science papers when experts interpreted the main findings of the studies into more accessible languages.

Although we and Hoogeveen et al.~\cite{hoogeveen2020laypeople} both focused on the prediction performance of laypeople, our experiments had four main differences from theirs, which may explain the prediction performance drop we found. 

\begin{itemize}
\item \textbf{Expert intervention.}
For each study, Hoogeveen et al. presented participants with a short description of the research question, the operationalization, and the key finding of the given study and then asked participants about the replication probability. These materials were composed and rephrased by experts to be comprehensible to laypeople. Thus, the participants might be more clear about the main findings of the studies. In fact, 72\% of participants indicated that they understood the descriptions of all the 27 studies used in Hoogeveen et al.'s experiments. In contrast, to reduce expert participation, for each study, we presented participants with the raw abstract and one experiment section directly truncated from the published paper of the study. This increased the cognitive burden of laypeople and raised the difficulty of understanding the studies' main findings. In our experiments, only 26\% of the responses indicated that they clearly understood the given abstract, and 41\% indicated that they had a general idea. This accessibility issue might further introduce vagueness in making their predictions about the replicability of the studies.

\item \textbf{Study set.} 
Hoogeveen et al. selected 27 studies from SSRP and ML2 replication projects. 
These two projects had a higher replication rate overall, 62\% for SSRP and 50\% for ML2, and the selected studies had a replication rate of 52\%. In contrast, we selected 89 studies from RPP due to its larger sample size (97 studies in RPP vs. 21 for SSRP and 24 for ML2).
However, the studies in RPP had a much lower replication rate, 37.5\% overall and 40\% for our 89 studies. This low replication rate creates a disadvantage, as people (both researcher forecasters and laypeople) tend to overestimate the replication rate. 
Moreover, the replicability of studies in RPP was more difficult to predict than SSRP and ML2. To see this point, the prediction accuracy score of researcher forecasters via surveys was 0.58 on RPP~\cite{dreber2015using}, compared to 0.86 on SSRP~\cite{camerer2018evaluating} and 0.67 on ML2~\cite{forsell2019predicting}. These features of RPP may partially explain why laypeople’s replication predictions had a salient overestimation and limited predictive power in our experiments.

\item \textbf{Participant population.} 
In Hoogeveen et al.'s experiments, most participants (54\%) were first-year students at the University
of Amsterdam, 32\% were Amazon Mturk workers, and the rest were recruited via social media. In contrast, all of our participants were Amazon Mturk workers. We conjectured that a student admitted by the psychology major of the world's renowned universities might have better skills in reading psychology papers and conducting related reasoning than an average Amazon Mturk worker. This advantage might contribute to more informative replication predictions in Hoogeveen et al.'s experiments.

\item \textbf{Presentation of material.}
Instead of presenting an interpreted description of the studies in Hoogeveen et al.'s experiments, we presented the participants with the raw material truncated from the published papers, which might create an impression to laypeople that these studies were designed and conducted thoughtfully and with rigorous examinations, potentially driving laypeople to make a higher replication prediction. This might also contribute to the salient overestimation of the replication probability. 

\end{itemize}

Given our results, we still think that there is a potential to rely on laypeople to make predictions about the replicability of social science studies, because we did find that the laypeople were willing to devote time and effort to complete our tasks. They did show an understanding of the materials to some extent and provided self-contained answers with good faith. In particular, they correctly identified the number of participants in the given study and formed sensible consensuses in selecting the main sentences of the materials. However, we also found that the information we elicited in the experiments from laypeople was not accurate in directly predicting the replicability of the studies.

On the one hand, there are efforts to use machine learning for a broader range of assessment tasks of published scientific studies~\cite{alipourfard2021systematizing,teja2021extraction,wu2021predicting}. We believe that the results from this study will be useful for generating training data, in particular for machine learning systems that do not train complex tasks end-to-end, but rather dissect complex tasks into simple tasks (e.g., such as extracting sample size, p-value, and the number of hypotheses tested) and train compositionally.

On the other hand, for predicting the replicability of social science studies, we do believe some adjustments to the information elicitation procedure might be necessary to increase the chance of success and will be exciting for future research. For example, we can provide participants with more refined material such as the main claim sentences and the main result sentences to reduce participants' cognitive burden and mitigate the vagueness in identifying the main findings. Moreover, some probabilistic training and replication judgment training may be required to help participants carry out necessary reasoning and reduce the overestimate bias. Furthermore, iterative and cooperative elicitation processes can be explored besides using one-shot surveys. For example, we can ask laypeople to articulate the main claims into more comprehensible languages iteratively and then make predictions.

\section*{Acknowledgments}
Yiling Chen, Michael Gordon, Thomas Pfeiffe and Juntao Wang thank the Fetzer Franklin Fund's Metascience grant program. Anna Dreber thanks the Jan Wallander and Tom Hedelius Foundation, the Knut and Alice Wallenberg Foundation, the Marianne and Marcus Wallenberg Foundation and the Austrian Science Fund (FWF, SFB F63). Anna Dreber and Magnus thank Riksbankens Jubileumsfond (grant P21-0168).

\bibliographystyle{plainnat}
\bibliography{references}

\begin{thebibliography}{28}
\providecommand{\natexlab}[1]{#1}
\providecommand{\url}[1]{\texttt{#1}}
\expandafter\ifx\csname urlstyle\endcsname\relax
  \providecommand{\doi}[1]{doi: #1}\else
  \providecommand{\doi}{doi: \begingroup \urlstyle{rm}\Url}\fi

\bibitem[Alipourfard et~al.(2021)Alipourfard, Arendt, Benjamin, Benkler,
  Bishop, Burstein, Bush, Caverlee, Chen, Clark,
  et~al.]{alipourfard2021systematizing}
Nazanin Alipourfard, Beatrix Arendt, Daniel~M Benjamin, Noam Benkler, Michael
  Bishop, Mark Burstein, Martin Bush, James Caverlee, Yiling Chen, Chae Clark,
  et~al.
\newblock Systematizing confidence in open research and evidence (score).
\newblock 2021.

\bibitem[Altmejd et~al.(2019)Altmejd, Dreber, Forsell, Huber, Imai,
  Johannesson, Kirchler, Nave, and Camerer]{altmejd2019predicting}
Adam Altmejd, Anna Dreber, Eskil Forsell, Juergen Huber, Taisuke Imai, Magnus
  Johannesson, Michael Kirchler, Gideon Nave, and Colin Camerer.
\newblock Predicting the replicability of social science lab experiments.
\newblock \emph{PloS one}, 14\penalty0 (12):\penalty0 e0225826, 2019.

\bibitem[Baker(2016)]{baker20161}
Monya Baker.
\newblock 1,500 scientists lift the lid on reproducibility.
\newblock \emph{Nature}, 533\penalty0 (7604), 2016.

\bibitem[Camerer et~al.(2016)Camerer, Dreber, Forsell, Ho, Huber, Johannesson,
  Kirchler, Almenberg, Altmejd, Chan, et~al.]{camerer2016evaluating}
Colin~F Camerer, Anna Dreber, Eskil Forsell, Teck-Hua Ho, J{\"u}rgen Huber,
  Magnus Johannesson, Michael Kirchler, Johan Almenberg, Adam Altmejd, Taizan
  Chan, et~al.
\newblock Evaluating replicability of laboratory experiments in economics.
\newblock \emph{Science}, 351\penalty0 (6280):\penalty0 1433--1436, 2016.

\bibitem[Camerer et~al.(2018)Camerer, Dreber, Holzmeister, Ho, Huber,
  Johannesson, Kirchler, Nave, Nosek, Pfeiffer, et~al.]{camerer2018evaluating}
Colin~F Camerer, Anna Dreber, Felix Holzmeister, Teck-Hua Ho, J{\"u}rgen Huber,
  Magnus Johannesson, Michael Kirchler, Gideon Nave, Brian~A Nosek, Thomas
  Pfeiffer, et~al.
\newblock Evaluating the replicability of social science experiments in nature
  and science between 2010 and 2015.
\newblock \emph{Nature Human Behaviour}, 2\penalty0 (9):\penalty0 637--644,
  2018.

\bibitem[Christensen and Miguel(2016)]{christensen2016transparency}
G~Christensen and E~Miguel.
\newblock Transparency, reproducibility, and the credibility of economics
  research. pt nber wp 22989. forthcoming in the.
\newblock \emph{Journal of Economic Literature}, 2016.

\bibitem[Collaboration(2015)]{open2015estimating}
Open~Science Collaboration.
\newblock Estimating the reproducibility of psychological science.
\newblock \emph{Science}, 349\penalty0 (6251):\penalty0 aac4716, 2015.

\bibitem[Davis and Goadrich(2006)]{davis2006relationship}
Jesse Davis and Mark Goadrich.
\newblock The relationship between precision-recall and roc curves.
\newblock In \emph{Proceedings of the 23rd international conference on Machine
  learning}, pages 233--240, 2006.

\bibitem[Dreber et~al.(2015)Dreber, Pfeiffer, Almenberg, Isaksson, Wilson,
  Chen, Nosek, and Johannesson]{dreber2015using}
Anna Dreber, Thomas Pfeiffer, Johan Almenberg, Siri Isaksson, Brad Wilson,
  Yiling Chen, Brian~A Nosek, and Magnus Johannesson.
\newblock Using prediction markets to estimate the reproducibility of
  scientific research.
\newblock \emph{Proceedings of the National Academy of Sciences}, 112\penalty0
  (50):\penalty0 15343--15347, 2015.

\bibitem[Fanelli(2018)]{fanelli2018opinion}
Daniele Fanelli.
\newblock Opinion: Is science really facing a reproducibility crisis, and do we
  need it to?
\newblock \emph{Proceedings of the National Academy of Sciences}, 115\penalty0
  (11):\penalty0 2628--2631, 2018.

\bibitem[Fidler and Wilcox(2021)]{sep-scientific-reproducibility}
Fiona Fidler and John Wilcox.
\newblock {Reproducibility of Scientific Results}.
\newblock In Edward~N. Zalta, editor, \emph{The {Stanford} Encyclopedia of
  Philosophy}. Metaphysics Research Lab, Stanford University, {S}ummer 2021
  edition, 2021.

\bibitem[Forsell et~al.(2019)Forsell, Viganola, Pfeiffer, Almenberg, Wilson,
  Chen, Nosek, Johannesson, and Dreber]{forsell2019predicting}
Eskil Forsell, Domenico Viganola, Thomas Pfeiffer, Johan Almenberg, Brad
  Wilson, Yiling Chen, Brian~A Nosek, Magnus Johannesson, and Anna Dreber.
\newblock Predicting replication outcomes in the many labs 2 study.
\newblock \emph{Journal of Economic Psychology}, 75:\penalty0 102117, 2019.

\bibitem[Galton(1907)]{galton1907vox}
Francis Galton.
\newblock Vox populi, 1907.

\bibitem[Goldstein et~al.(2014)Goldstein, McAfee, and
  Suri]{goldstein2014wisdom}
Daniel~G Goldstein, Randolph~Preston McAfee, and Siddharth Suri.
\newblock The wisdom of smaller, smarter crowds.
\newblock In \emph{Proceedings of the fifteenth ACM conference on Economics and
  computation}, pages 471--488, 2014.

\bibitem[Gordon et~al.(2020)Gordon, Viganola, Bishop, Chen, Dreber, Goldfedder,
  Holzmeister, Johannesson, Liu, Twardy, et~al.]{gordon2020replication}
Michael Gordon, Domenico Viganola, Michael Bishop, Yiling Chen, Anna Dreber,
  Brandon Goldfedder, Felix Holzmeister, Magnus Johannesson, Yang Liu, Charles
  Twardy, et~al.
\newblock Are replication rates the same across academic fields? community
  forecasts from the darpa score programme.
\newblock \emph{Royal Society open science}, 2020.

\bibitem[Gordon et~al.(2021)Gordon, Viganola, Dreber, Johannesson, and
  Pfeiffer]{gordon2021predicting}
Michael Gordon, Domenico Viganola, Anna Dreber, Magnus Johannesson, and Thomas
  Pfeiffer.
\newblock Predicting replicability—analysis of survey and prediction market
  data from large-scale forecasting projects.
\newblock \emph{Plos one}, 16\penalty0 (4):\penalty0 e0248780, 2021.

\bibitem[Hoogeveen et~al.(2020)Hoogeveen, Sarafoglou, and
  Wagenmakers]{hoogeveen2020laypeople}
Suzanne Hoogeveen, Alexandra Sarafoglou, and Eric-Jan Wagenmakers.
\newblock Laypeople can predict which social-science studies will be replicated
  successfully.
\newblock \emph{Advances in Methods and Practices in Psychological Science},
  3\penalty0 (3):\penalty0 267--285, 2020.

\bibitem[Ioannidis and Doucouliagos(2013)]{ioannidis2013s}
John Ioannidis and Chris Doucouliagos.
\newblock What's to know about the credibility of empirical economics?
\newblock \emph{Journal of Economic Surveys}, 27\penalty0 (5):\penalty0
  997--1004, 2013.

\bibitem[Ioannidis(2005)]{ioannidis2005most}
John~PA Ioannidis.
\newblock Why most published research findings are false.
\newblock \emph{PLoS medicine}, 2\penalty0 (8):\penalty0 e124, 2005.

\bibitem[Klein et~al.(2018)Klein, Vianello, Hasselman, Adams, Adams~Jr, Alper,
  Aveyard, Axt, Babalola, Bahn{\'\i}k, et~al.]{klein2018many}
Richard~A Klein, Michelangelo Vianello, Fred Hasselman, Byron~G Adams,
  Reginald~B Adams~Jr, Sinan Alper, Mark Aveyard, Jordan~R Axt, Mayowa~T
  Babalola, {\v{S}}t{\v{e}}p{\'a}n Bahn{\'\i}k, et~al.
\newblock Many labs 2: Investigating variation in replicability across samples
  and settings.
\newblock \emph{Advances in Methods and Practices in Psychological Science},
  1\penalty0 (4):\penalty0 443--490, 2018.

\bibitem[Maniadis et~al.(2014)Maniadis, Tufano, and List]{maniadis2014one}
Zacharias Maniadis, Fabio Tufano, and John~A List.
\newblock One swallow doesn't make a summer: New evidence on anchoring effects.
\newblock \emph{American Economic Review}, 104\penalty0 (1):\penalty0 277--90,
  2014.

\bibitem[Mellers et~al.(2014)Mellers, Ungar, Baron, Ramos, Gurcay, Fincher,
  Scott, Moore, Atanasov, Swift, et~al.]{mellers2014psychological}
Barbara Mellers, Lyle Ungar, Jonathan Baron, Jaime Ramos, Burcu Gurcay, Katrina
  Fincher, Sydney~E Scott, Don Moore, Pavel Atanasov, Samuel~A Swift, et~al.
\newblock Psychological strategies for winning a geopolitical forecasting
  tournament.
\newblock \emph{Psychological science}, 25\penalty0 (5):\penalty0 1106--1115,
  2014.

\bibitem[Prelec et~al.(2017)Prelec, Seung, and McCoy]{prelec2017solution}
Dra{\v{z}}en Prelec, H~Sebastian Seung, and John McCoy.
\newblock A solution to the single-question crowd wisdom problem.
\newblock \emph{Nature}, 541\penalty0 (7638):\penalty0 532--535, 2017.

\bibitem[Salsabil et~al.(2018)Salsabil, Wu, Choudhury, Ingram, Fox, Rajtmajer,
  and Giles]{salsabil2018study}
Lamia Salsabil, Jian Wu, Muntabir~Hasan Choudhury, William~A Ingram, Edward~A
  Fox, Sarah~J Rajtmajer, and C~Lee Giles.
\newblock A study of computational reproducibility using urls linking to open
  access datasets and software.
\newblock 2018.

\bibitem[Schmidt(2016)]{schmidt2016shall}
Stefan Schmidt.
\newblock Shall we really do it again? the powerful concept of replication is
  neglected in the social sciences.
\newblock 2016.

\bibitem[Teja~Lanka et~al.(2021)Teja~Lanka, Rajtmajer, Wu, and
  Giles]{teja2021extraction}
Sree~Sai Teja~Lanka, Sarah Rajtmajer, Jian Wu, and C~Lee Giles.
\newblock Extraction and evaluation of statistical information from social and
  behavioral science papers.
\newblock In \emph{Companion Proceedings of the Web Conference 2021}, pages
  426--430, 2021.

\bibitem[Wu et~al.(2021)Wu, Nivargi, Lanka, Menon, Modukuri, Nakshatri, Wei,
  Wang, Caverlee, Rajtmajer, et~al.]{wu2021predicting}
Jian Wu, Rajal Nivargi, Sree Sai~Teja Lanka, Arjun~Manoj Menon, Sai~Ajay
  Modukuri, Nishanth Nakshatri, Xin Wei, Zhuoer Wang, James Caverlee, Sarah~M
  Rajtmajer, et~al.
\newblock Predicting the reproducibility of social and behavioral science
  papers using supervised learning models.
\newblock \emph{arXiv preprint arXiv:2104.04580}, 2021.

\bibitem[Yang et~al.(2020)Yang, Youyou, and Uzzi]{yang2020estimating}
Yang Yang, Wu~Youyou, and Brian Uzzi.
\newblock Estimating the deep replicability of scientific findings using human
  and artificial intelligence.
\newblock \emph{Proceedings of the National Academy of Sciences}, 117\penalty0
  (20):\penalty0 10762--10768, 2020.

\end{thebibliography}

\end{spacing}
\end{document}